%% file: main.tex
\def\year{2018}\relax
\documentclass[letterpaper]{article} 
\usepackage{aaai18}  
\usepackage{times}  
\usepackage{helvet}  
\usepackage{balance} 

\usepackage{booktabs} 
\usepackage{balance}

\usepackage{courier}  
\usepackage{url}  
\usepackage{graphicx}  
\begin{document}
%
\title{Popularity-Aware Item Weighting for Long-Tail Recommendation}
\author{Himan Abdollahpouri, Robin Burke, Bamshad Mobasher\\
Web Intelligence Lab\\
College of Computing and Digital Media\\
DePaul University, Chicago, USA\\
habdolla@depaul.edu, rburke@cs.depaul.edu, mobasher@depaul.edu
}
\maketitle
\begin{abstract}
Many recommender systems suffer from the popularity bias problem: popular items are being recommended frequently while less popular, niche products, are recommended rarely if not at all. However, those ignored products are exactly the products that businesses need to find customers for and their recommendations would be more beneficial. In this paper, we examine an item weighting approach to improve long-tail recommendation. Our approach works as a simple yet powerful add-on to existing recommendation algorithms for making a tunable trade-off between accuracy and long-tail coverage. 
\end{abstract}

\input{body-v1}
\balance
\bibliographystyle{aaai}
\bibliography{main.bib}

\end{document}

%% file: body-v1.tex
\section{Introduction}
\noindent 
Recommender systems have an important role in e-commerce and information sites, helping users find new items. One obstacle to effectiveness of recommenders is in the problem of popularity bias: collaborative filtering recommenders typically emphasize popular items (those with more ratings) much more than other ``long-tail'' items~\cite{longtailrecsys}. Although popular items are often good recommendations, they are also likely to be well-known, so delivering only on popular items will not enhance new item discovery.

\begin{figure}[tbh]
    \centering
    \includegraphics[height=2.5in, width=3.5in]{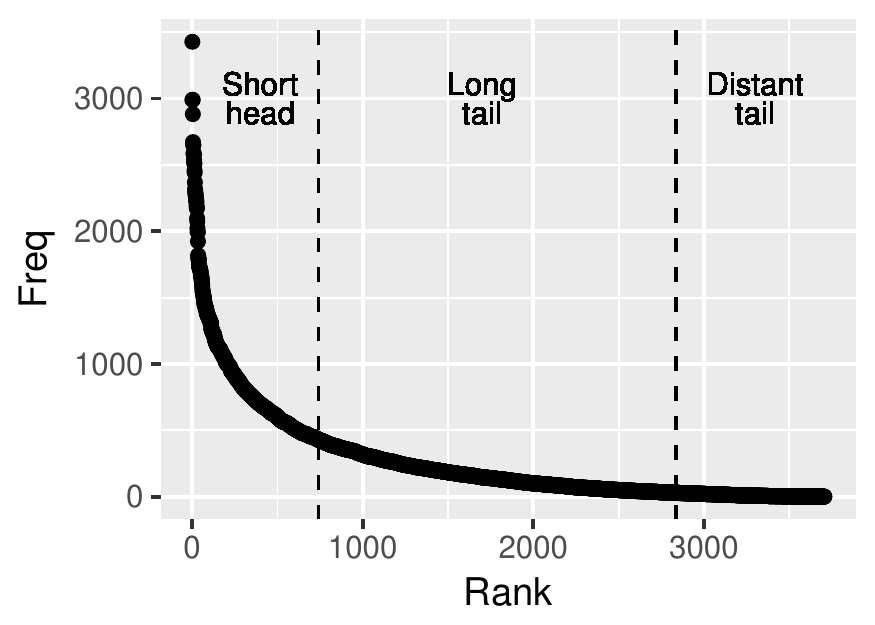}
    \caption{The long-tail of item popularity. }
    \label{fig:lt-effect}
\end{figure}

Figure~\ref{fig:lt-effect} illustrates the long-tail phenomenon in the well-known MovieLens 1M dataset~\cite{movielens}. The $y$ axis represents the number of ratings per item and the $x$ axis shows the product rank. The first vertical line separates the top 20\% of items by popularity -- these items cumulatively have many more ratings than the 80\%  tail items to the right. These ``short head'' items are the very popular blockbuster movies that garner much more viewer attention. Similar distributions can be found in books, music, and other consumer taste domains. 

The second vertical line divides the tail of the distribution into two parts. The first part we call the \textit{long tail}: these items are accessible to collaborative recommendation, even though recommendation algorithms often do not produce them. Beyond this point, items receive so few ratings that meaningful cross-user comparison of their ratings becomes noisy and unreliable. For these items, the \textit{distant tail} or cold-start items, content-based and hybrid recommendation techniques must be employed. Our work in this paper is concerned with collaborative recommendation and therefore focuses on the long tail.

Long-tail recommendation can be understood as an example of a more general phenomenon of non-uniform item preference within a recommender system. The system may have preferences over the items in the catalog and as a result, may have the goal of promoting certain items and demoting others. Recommendation approaches that are sensitive to the value of items (or users) to the systems are known as \textit{value-aware} recommendation \cite{amatriain2016past}. 

We present a general and flexible framework for value-aware recommendation and, in particular, long-tail recommendation built on the top of the standard recommendation algorithms. When applied to long-tail items, this approach enables the system designer to tune the application to achieve a particular trade-off between accuracy and the inclusion of long-tail items. However, the framework is sufficiently general that it can be used to control recommendation bias for or against any group of items. 

\subsection{Related Work}

Recommending serendipitous items from the long tail is generally considered to be a key function of recommendation \cite{shani2011evaluating,anderson2006long}, as these are items that users are less likely to know about. Brynjolfsson and his colleagues showed that 30-40\% of Amazon book sales are represented by titles that would not normally be found in brick-and-mortar stores~\cite{brynjolfsson2006niches}. Access to long-tail items is therefore a strong driver for e-commerce growth.

Long-tail items are also important for generating a fuller understanding of users' preferences. Systems that use active learning to explore each user's profile will typically need to present more long tail items because these are the ones that the user is less likely to have rated, and where user's preferences are more likely to be diverse~\cite{nguyen2014exploring,resnick2013bursting,abdollahpouri2017towards}. 

Finally, long-tail recommendation can also be understood as a social good. A market that suffers from popularity bias will lack opportunities to discover more obscure products and will be, by definition, dominated by a few large brands or well-known artists~\cite{celma2008hits}. Such a market will be more homogeneous and offer fewer opportunities for innovation and creativity.

The idea of the long-tail of item popularity and its impact on recommendation quality has been explored by some researchers \cite{brynjolfsson2006niches,longtailrecsys}. In those works, authors tried to improve the performance of the recommender system in terms of accuracy and precision, given the long-tail in the ratings. Our work, instead, focuses on reducing popularity bias and balancing the presentation of items across the popularity distribution.  

There are also some works that try to improve long tail recommendations such as \cite{abdollahpouri2017controlling} where authors use regularization on the top of a learning-to-rank algorithm to balance the proportion of long-tail and short-head items in the list. That work had two key limitations that are avoided in this work. First, the distinction between long-tail and short-head items was a hard binary distinction, whereas in this work, we use a graduated, weighted approach to the long-tail preference. Second, the regularization approach is restricted to factorization models where the long-tail preference can be encoded in terms of the latent factors. Because our approach involves post-processing the recommendation output, it can be applied to any collaborative recommendation algorithm.

\section{System-level preferences}
In some situations, we need to incorporate system concerns about the products and users into the recommendation generation. For instance, the system might want to promote certain products more often than the others for a variety of reasons such as profitability, fairness and other financial or non-financial purposes~\cite{moveRecProfitMax,DBLP:conf/um/BurkeAMG16}. Moreover, the system might want certain products to be recommended more often to certain users. For example, some users might have premium membership in the system and the system might give higher priorities to those users. What matters here is, for any reason, the system might prefer the recommendation of some items over others. Current recommendation algorithms, in their existing form, cannot handle this type of concerns directly and often businesses have to apply extra processing to recommender system outputs to get the desired results. In other words, there is no consistent and standard way to handle extra system level preferences. 

\begin{figure}[bh]
    \centering
    \includegraphics[width=3.5in]{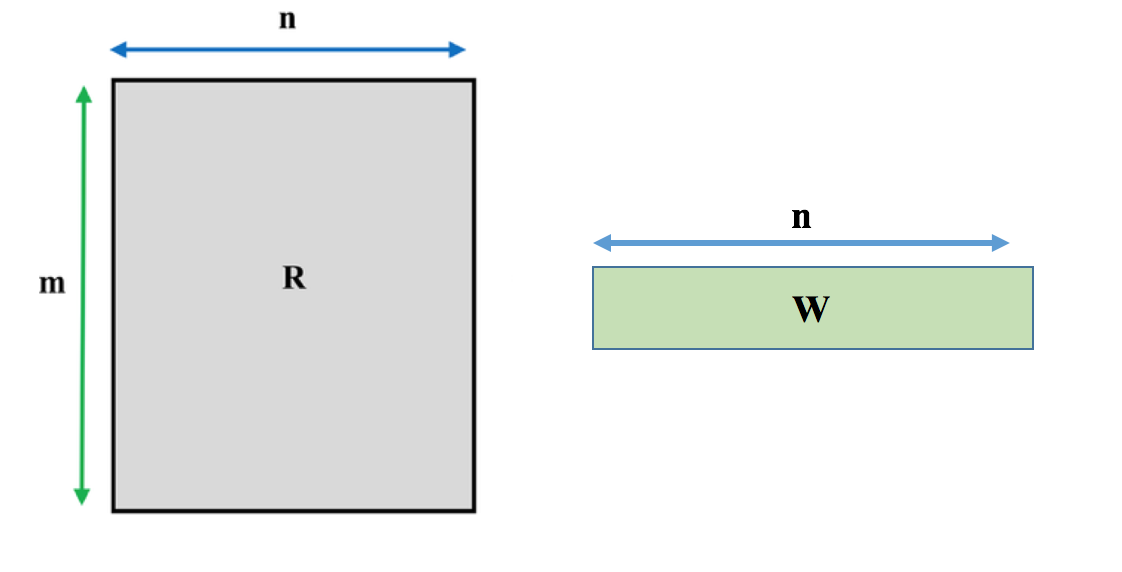}
    \caption{Item preference vector W along with the user rating preference matrix R }
    \label{fig:formulation}
\end{figure}

\subsection{Item weighting}

Item weighting is a simple yet extendable approach that can be used with any recommendation algorithm to handle variety of different system level preferences. The idea of item weighting is to systematically incorporate system preferences over different items and try to generate the recommendations by taking into account both user and system preferences. In addition to a matrix of ratings preferences $R$ in which each entry is represented as $r_{ui}$, denoting the rating given by user $u$ to item $i$, we define an extra vector, $W$ in which the system's preferences over different items are expressed. See Figure~\ref{fig:formulation}. That is, each entry $w_i$ in vector $W$ represents the preference level of the system towards the corresponding item.

We apply this idea to long-tail promotion, which is an example of such a system-level preference. For this purpose, we need a weighting scheme that gives higher weight to long-tail items and lesser weight to the popular items, to counterbalance the bias that their popularity induces. Our choice for the $w_i$ values in this paper is as follows:

\begin{equation}\label{eq:weight}
    w_i=\frac{1}{log(\rho(i))}
\end{equation}
where $\rho(i)$ denotes the number of times that item $i$ has been rated. The log function in the denominator limits the range of this value so that popular items are not highly penalized. In future work, we plan to explore other related weighting functions.

\subsection{Value-aware ranking score}
In some formulations of the recommendation problem, the task is to predict the rating that a user would provide to previously unseen items. If the desired output for a recommender system is a ranked list of items for each user, predicted ratings can be used to sort and select to create such a list. However, predicted ratings are not necessary for ranking and in this work, we aim instead to produce a ranking score that incorporates both the predicted rating and the system preference.


Let $\hat{r}_{ui}$ be the predicted rating for a user-item pair from some recommendation algorithm. Then the value-aware ranking score for that pair would be:

\begin{equation}\label{eq:ranking}
    \Upsilon_{ui}= (1-\alpha) \hat{r}_{ui} + \alpha w_i
\end{equation}
where $\alpha$ is the adjusting weight that controls the level of importance of each of the two parts in Equation~\ref{eq:ranking}. Smaller values for $\alpha$ mean more weight for user's interest as expressed in the user profile and higher values for $\alpha$ shift the balance towards the system's preferences as expressed in the $W$ vector. 

The $\hat{r}_{ui}$ values can be calculated using any standard collaborative filtering algorithms such as matrix factorization, user-based or item-based KNN etc. \cite{su2009survey}. It could also be the ranking score calculated by learning to rank algorithms for any given user and item where higher scores correspond to items on the top of the list \cite{pessiot2007learning}.  

Once $\Upsilon_{ui}$ is calculated for each pair, the top-k items are selected to be recommended to user $u$. The logic behind this formulation is, since the predicted scores are the combination of both system-level and user-level preferences, the final recommendations give higher utility to the system, given that $\alpha$ is set to an appropriate value.

\section{Evaluation}
As we discussed above, the system, for any reason, might prefer to promote certain items  more often than others. We denoted that preference as a $1 \times N$ vector representing preference values of system over any of the items. Since we are interested in incorporating extra concerns in our recommendations (in this case promoting long tail products), the standard evaluation metrics such as precision etc. are not able to capture that aspect of performance as they are designed to capture the accuracy of the recommendations. Therefore, in this paper, we define two extra metrics that are able to capture the success of a recommender system in terms of long-tail promotion. The first metric is called \textit{Recommendations Popularity (RP)} which is the average total number of times each of the recommended items were rated in the training set and is defined as follows:

\begin{equation}\label{eq:rp}
     RP=\frac{\sum_{u \in U_{t}} \sum_{i \in L(u)} \rho(i)}{|U_t|\times K}
\end{equation}
where $|U_{t}|$ is the number of users in the test set and $\rho(i)$ denotes the number of times that item $i$ has been rated in the train set. $K$ is the length of the recommendation list (in our experiments it is 10) and $L(u)$ is the recommendation list for user $u$. \textit{RP} shows, on average, how popular the recommended items are. Lower values for \textit{RP} represent better long-tail promotion. 
A different but related metric of success is the Average Percentage of Long-tail items (APL) in the recommendation lists, which we define as follows: 

\begin{equation}\label{eq:apl}
     APL=\frac{1}{|U_{t}|}\sum_{u \in U_{t}} \frac{|\{i, i \in (L(u) \cap  \Phi)\} |}{|L(u)|}
\end{equation}
where  $\Phi$ is the set of long-tail items (i.e. items which are between the two vertical lines in Figure~\ref{fig:lt-effect}). This measure tells us on average what percentage of items in users' recommendation lists belongs to the long-tail set.

A third measure of interest in long-tail recommendation is \textit{long-tail coverage}. To understand the importance of this measure, consider a recommender that selected a particular small set of long-tail items and included them in every recommendation list. This recommender might have good scores on the RP and APL metrics, but it is not achieving the goal of exposing a variety of long tail items to users. Long-tail catalog coverage measures the fraction of unique long tail items across the entire set of recommendation lists for all users relative to the long-tail set. More formally, $LCC$ is defined as:

\begin{equation}\label{eq:coverage}
LCC = \frac{|(\bigcup_{u \in U_t} L_u) \cap \Phi|}{|\Phi|}
\end{equation}

We also evaluate the accuracy of the recommendations using \textit{precision@10}.

\section{Methodology}

\noindent We used two data sets in our experiments. The first  is the well-known Movielens 1M data set that contains 1,000,209 anonymous ratings of approximately 3,900 movies made by 6,040 MovieLens users who joined MovieLens in 2000~\cite{movielens}. The second data set is the Epinions data set, which is gathered from a consumers opinion site where users can review items (such as cars, books, movies, and software) and  assign them numeric ratings in the range 1 (min) to 5 (max) \cite{massa2007trust}. This data set has total number of 664,824 ratings given by 40,163 users to 139,736 items. In Movielens, each user has a minimum of 20 ratings but in Epinions, there are many users with only a single rated item. 

Following the example of \cite{abdollahpouri2017controlling}, we removed users who had fewer than 30 ratings, as these users are unlikely to have long-tail items in their profiles. The retained users were those likely to have rated enough long-tail items so that our objective could be evaluated in a train / test scenario. We also removed distant long-tail items from each data set, using a limit of 30 ratings. This has a very significant impact on the data set, as there is a huge distant part of the tail containing many cold-start products with only one rating.

After the removal of short-profile users and distant long tail items, the Movielens data set has 5,289 users who rated 2,836 movies with a total number of 972,471 ratings, a reduction of about 3\%. Applying the same criteria to the Epinions data set decreases the data to 157,887 ratings given by 5,383 users to 3,423 items, a reduction of more than 75\%.

We used Librec \cite{guo2015librec}, an open source Java framework for recommender system implementation, modified by the inclusion of our item weighting framework. We used a random split of 80\% of the data for training and the remaining 20\% for testing. For each user in the test set, we recommended a list of 10 items using four different algorithms: Bayesian Personalized Ranking (BPR) \cite{rendle2009bpr}, Alternating Least Squares for personalized ranking (ALS) \cite{takacs2012alternating}, and two simple baseline algorithms called most popular recommendation (Pop) and random guess recommendation (Rand). These last two are extremes with respect to long-tail recommendation: a random recommender is insensitive to popularity and will recommend items across the catalog; a  popularity-based recommender will ignore long-tail items entirely and recommend only a very small number of popular items to everyone.

We evaluated the results using four metrics: precision@10, to capture the accuracy of the personalized recommendations; the APL measure from Equation~\ref{eq:apl} indicating proportion of long-tail items in the average recommendation list; recommendation popularity (RP) as defined in equation~\ref{eq:rp}, the average number of times each recommended item in the list has been rated; and, long-tail catalog coverage (LCC) as defined in equation~\ref{eq:coverage}.

\section{Results}

\begin{figure*}[ht]
    \centering
    \includegraphics[height=2.6in, width=6.5in]{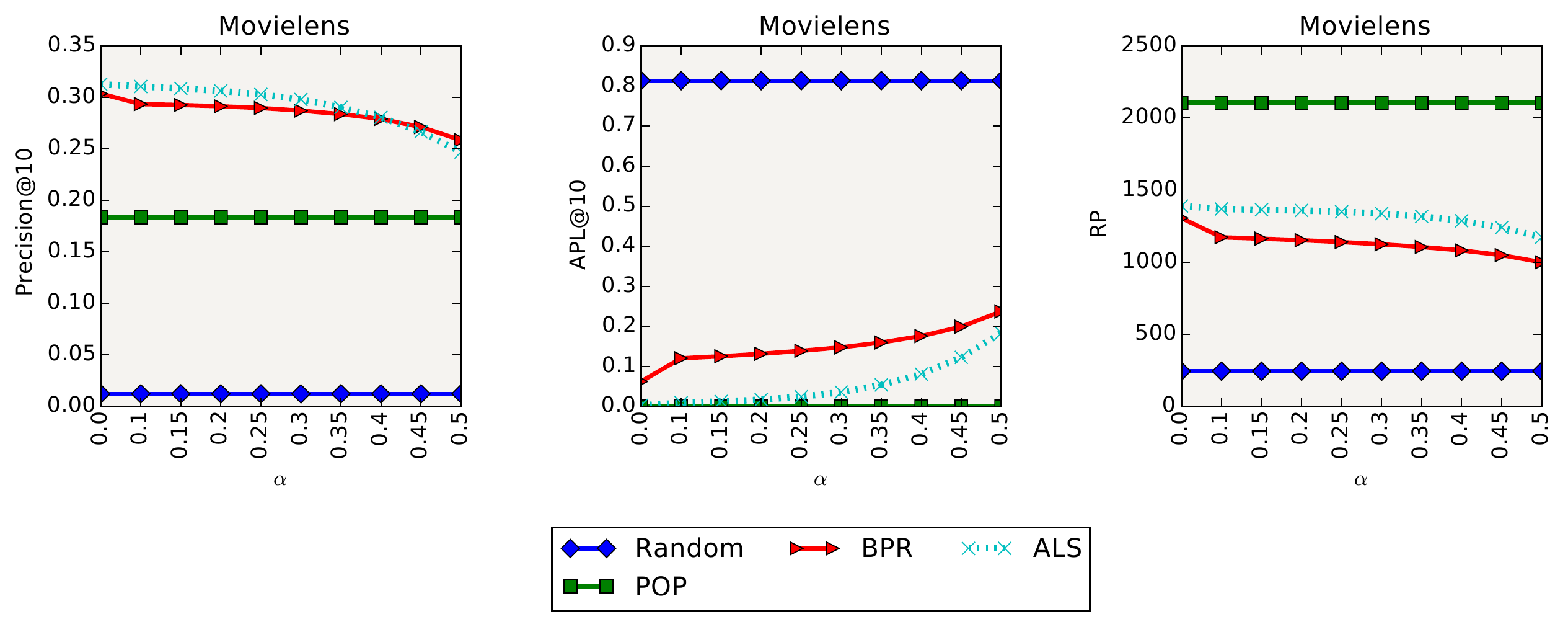}
    \caption{The Precision@10, APL@10, and RP  (Movielens)}
    \label{fig:ml-results}
\end{figure*}

\begin{figure*}[ht]
    \centering
    \includegraphics[height=2.6in, width=6.5in]{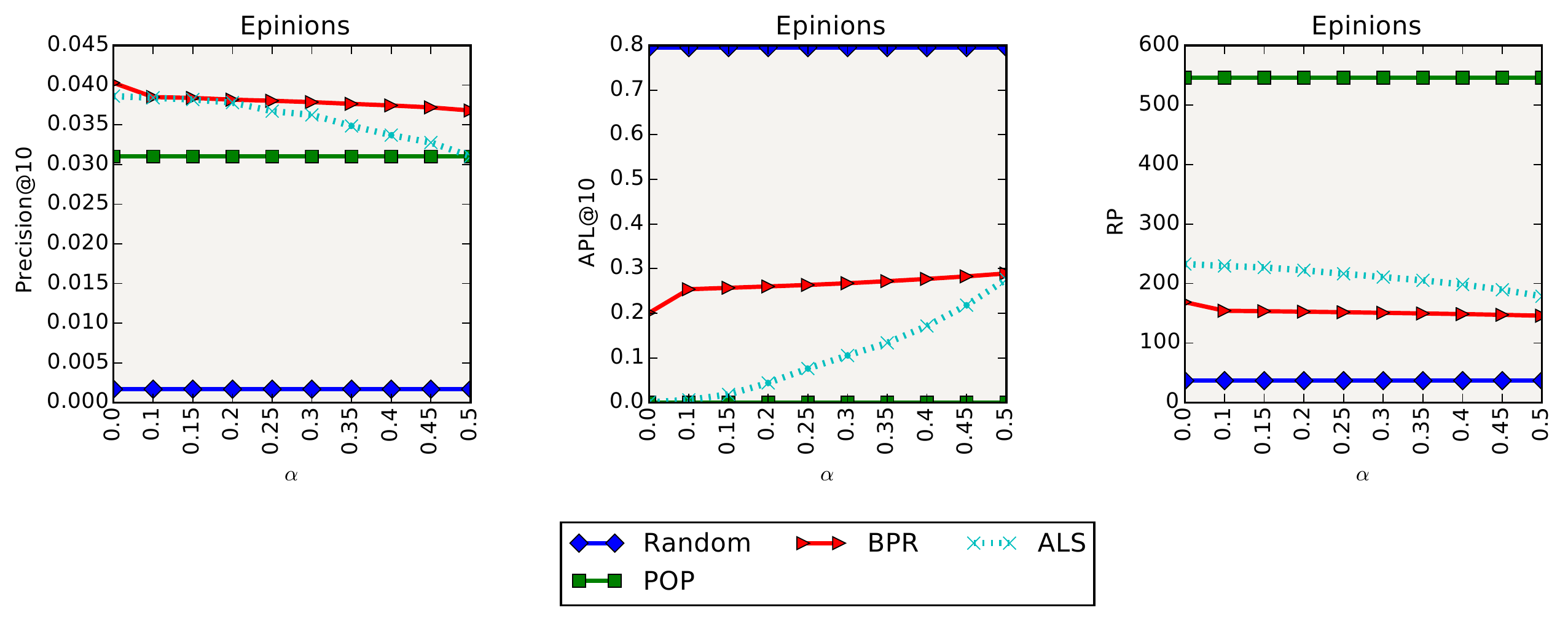}
    \caption{The Precision@10, APL@10, and RP  (Epinions)}
    \label{fig:ep-results}
\end{figure*}

Figure~\ref{fig:ml-results} shows the results of the experiment on the MovieLens 1M data set. The three visualizations indicate that the item weighting scheme works as designed. The average number of long-tail items increases as $\alpha$ increases, placing more weight on the $w_i$ terms in the ranking score. At $\alpha = 0.5$, both BPR and ALS are including approximately 20\% long-tail items (or an average of two such items per 10-item list.) The recommendation popularity graph tells a similar story, with the average popularity of the recommended items declining over the same interval. At the same time, the precision for both algorithms declines about 16\% from around 0.3 to 0.25. 

The results for Epinions shown in Figure~\ref{fig:ep-results} are similar. This is a more difficult recommendation problem due to the nature of the datset so the precision is smaller. The APL score increases more slowly for the ALS algorithm, but eventually reaches the same level (around 0.3) at $\alpha = 0.5$. The RP results tell a similar story. Over both data sets, these results show that there is a range of $\alpha$ values over which long-tail performance can be increased with minimal loss to recommendation accuracy. 

An interesting twist to the story is provided by the data shown in Tables~\ref{tab1} and \ref{tab2}. We can see that total long-tail coverage for the ALS algorithm is only a fraction of that of BPR. Even in the baseline $\alpha = 0$ condition, BPR is returning 15 times as many long-tail items and this number increases over the $\alpha$ range. The ALS fraction barely increases, which means that it is (most likely) achieving its increased APL score by recommending the same items to more users. A similar pattern is seen in the Epinions data set. Here ALS does increase its coverage over the parameter range, but it is just a fraction of what BPR is able to achieve.

Clearly, an implementer interested in long-tail diversity would do best by using the BPR algorithm over ALS. It has better long-tail performance in its unmodified form and with a small loss in precision, both the long-tail contents of recommendation lists and the long-tail coverage can be improved significantly.

One reason for this difference may have to do with the different objectives that each algorithm employs. BPR uses an objective that takes into account pairwise ordering relations. So, if a user prefers item $i$ over item $j$, the algorithm tries to learn latent factors such that this ranking is retained. The RankALS algorithm uses an objective that tries to ensure that the difference in rating values between items are learned by the factorization. This is a much more stringent requirement that just preserving the ordering. It appears that the strictness of this objective penalizes most long-tail items, which are more likely to have lower ratings, and only a small number of long-tail items get enough weight in the learned factors to be ranked highly on any recommendation list.

\begin{table}[tbp]
\centering
\caption{Total long-tail coverage (LCC)}{Movielens} \label{tab1} 
\begin{tabular}{r|r|r|r}
\hline 
Algorithm & $\alpha=0$ & $\alpha=0.25$ & $\alpha=0.5$ \\ \hline
BPR & 0.237 & 0.417 & 0.497 \\
ALS & 0.0135 &0.0184 &0.0143 \\\hline
\hline
\end{tabular}
\end{table}

\begin{table}[tbp]
\centering
\caption{Total long-tail coverage (LCC)}{Epinions} \label{tab2} 
\begin{tabular}{r|r|r|r}
\hline 
Algorithm & $\alpha=0$ & $\alpha=0.25$ & $\alpha=0.5$ \\ \hline
BPR & 0.540 & 0.637 & 0.657  \\
ALS & 0.003 & 0.009 & 0.008\\\hline
\hline
\end{tabular}
\end{table}

\section{Conclusion and future work}
Long-tail items are an important key to practical success for recommender systems. Since short-head items are likely to be well known to many users, the ability to recommend items outside of this band of popularity will determine if a recommender can introduce users to new products and experiences. Yet, it is well-known that recommendation algorithms have biases towards popular items.

In this paper, we presented an extensible framework that can be used on the top of any recommendation algorithm to account for extra system-level preferences over different items. As an example of system-level preference, we showed that it is possible to model the trade-off between long-tail catalog coverage and accuracy as a weighted hybrid score calculation for any pair of users and items. We also showed that, in contrast to BPR, the RankALS algorithm has very poor long-tail catalog coverage, a defect that cannot be remedied with an item-weighting approach.

In addition to the long-tail question that we address here, our value-aware item weighting framework can model many other system-level preferences that can be expressed as numerical or ordinal values. For example, in profit maximization, the weights in vector \textit{W} could be set to prices or profit margins and therefore, the recommendations could gain a better trade-off in terms of profitability and personalization.

One interesting area for future work could be taking into account users' individual differences when handling system-level preferences. For example, in our long-tail recommendation problem, some users are more open to receive long-tail recommendations compared to some others. Our model, in its current form, does not take this into account. 
Another future work would be using this model for multistakeholder recommendation where the system wants to make recommendations in the presence of different stakeholders providing the products \cite{umapHimanMS}. In those cases, the vector \textit{W} could be acquired based on the preferences of each stakeholder.  

%% file: main.bbl
\begin{thebibliography}{}

\bibitem[\protect\citeauthoryear{Abdollahpouri and
  Essinger}{2017}]{abdollahpouri2017towards}
Abdollahpouri, H., and Essinger, S.
\newblock 2017.
\newblock Towards effective exploration/exploitation in sequential music
  recommendation.

\bibitem[\protect\citeauthoryear{Abdollahpouri, Burke, and
  Mobasher}{2017a}]{abdollahpouri2017controlling}
Abdollahpouri, H.; Burke, R.; and Mobasher, B.
\newblock 2017a.
\newblock Controlling popularity bias in learning to rank recommendation.
\newblock In {\em Proceedings of the 11th ACM conference on Recommender
  systems. ACM, To appear}.

\bibitem[\protect\citeauthoryear{Abdollahpouri, Burke, and
  Mobasher}{2017b}]{umapHimanMS}
Abdollahpouri, H.; Burke, R.; and Mobasher, B.
\newblock 2017b.
\newblock Recommender systems as multi-stakeholder environments.
\newblock In {\em Proceedings of the 25th Conference on User Modeling,
  Adaptation and Personalization (UMAP2017)}.
\newblock ACM.

\bibitem[\protect\citeauthoryear{Amatriain and
  Basilico}{2016}]{amatriain2016past}
Amatriain, X., and Basilico, J.
\newblock 2016.
\newblock Past, present, and future of recommender systems: An industry
  perspective.
\newblock In {\em Proceedings of the 10th ACM Conference on Recommender
  Systems},  211--214.
\newblock ACM.

\bibitem[\protect\citeauthoryear{Anderson}{2006}]{anderson2006long}
Anderson, C.
\newblock 2006.
\newblock {\em The long tail: Why the future of business is selling more for
  less}.
\newblock Hyperion.

\bibitem[\protect\citeauthoryear{Azaria \bgroup et al\mbox.\egroup
  }{2013}]{moveRecProfitMax}
Azaria, A.; Hassidim, A.; Kraus, S.; Eshkol, A.; Weintraub, O.; and Netanely,
  I.
\newblock 2013.
\newblock Movie recommender system for profit maximization.
\newblock In {\em Proceedings of the 7th ACM conference on Recommender
  systems},  121--128.
\newblock ACM.

\bibitem[\protect\citeauthoryear{Brynjolfsson, Hu, and
  Smith}{2006}]{brynjolfsson2006niches}
Brynjolfsson, E.; Hu, Y.~J.; and Smith, M.~D.
\newblock 2006.
\newblock From niches to riches: Anatomy of the long tail.
\newblock {\em Sloan Management Review}  67--71.

\bibitem[\protect\citeauthoryear{Burke \bgroup et al\mbox.\egroup
  }{2016}]{DBLP:conf/um/BurkeAMG16}
Burke, R.~D.; Abdollahpouri, H.; Mobasher, B.; and Gupta, T.
\newblock 2016.
\newblock Towards multi-stakeholder utility evaluation of recommender systems.
\newblock In {\em Workshop on Surprise, Opposition, and Obstruction in Adaptive
  and Personalized Systems, UMAP 2016}.

\bibitem[\protect\citeauthoryear{Celma and Cano}{2008}]{celma2008hits}
Celma, {\`O}., and Cano, P.
\newblock 2008.
\newblock From hits to niches?: or how popular artists can bias music
  recommendation and discovery.
\newblock In {\em Proceedings of the 2nd KDD Workshop on Large-Scale
  Recommender Systems and the Netflix Prize Competition}, ~5.
\newblock ACM.

\bibitem[\protect\citeauthoryear{Guo \bgroup et al\mbox.\egroup
  }{2015}]{guo2015librec}
Guo, G.; Zhang, J.; Sun, Z.; and Yorke-Smith, N.
\newblock 2015.
\newblock Librec: A java library for recommender systems.
\newblock In {\em UMAP Workshops}.

\bibitem[\protect\citeauthoryear{Harper and Konstan}{2015}]{movielens}
Harper, F.~M., and Konstan, J.~A.
\newblock 2015.
\newblock The movielens datasets: History and context.
\newblock {\em ACM Transactions on Interactive Intelligent Systems (TiiS)}
  5(4):19.

\bibitem[\protect\citeauthoryear{Massa and Avesani}{2007}]{massa2007trust}
Massa, P., and Avesani, P.
\newblock 2007.
\newblock Trust-aware recommender systems.
\newblock In {\em Proceedings of the 2007 ACM conference on Recommender
  systems},  17--24.
\newblock ACM.

\bibitem[\protect\citeauthoryear{Nguyen \bgroup et al\mbox.\egroup
  }{2014}]{nguyen2014exploring}
Nguyen, T.~T.; Hui, P.-M.; Harper, F.~M.; Terveen, L.; and Konstan, J.~A.
\newblock 2014.
\newblock Exploring the filter bubble: the effect of using recommender systems
  on content diversity.
\newblock In {\em Proceedings of the 23rd international conference on World
  wide web},  677--686.
\newblock ACM.

\bibitem[\protect\citeauthoryear{Park and Tuzhilin}{2008}]{longtailrecsys}
Park, Y.-J., and Tuzhilin, A.
\newblock 2008.
\newblock The long tail of recommender systems and how to leverage it.
\newblock In {\em Proceedings of the 2008 ACM conference on Recommender
  systems},  11--18.
\newblock ACM.

\bibitem[\protect\citeauthoryear{Pessiot \bgroup et al\mbox.\egroup
  }{2007}]{pessiot2007learning}
Pessiot, J.-F.; Truong, V.; Usunier, N.; Amini, M.; and Gallinari, P.
\newblock 2007.
\newblock Learning to rank for collaborative filtering.

\bibitem[\protect\citeauthoryear{Rendle \bgroup et al\mbox.\egroup
  }{2009}]{rendle2009bpr}
Rendle, S.; Freudenthaler, C.; Gantner, Z.; and Schmidt-Thieme, L.
\newblock 2009.
\newblock Bpr: Bayesian personalized ranking from implicit feedback.
\newblock In {\em Proceedings of the twenty-fifth conference on uncertainty in
  artificial intelligence},  452--461.
\newblock AUAI Press.

\bibitem[\protect\citeauthoryear{Resnick \bgroup et al\mbox.\egroup
  }{2013}]{resnick2013bursting}
Resnick, P.; Garrett, R.~K.; Kriplean, T.; Munson, S.~A.; and Stroud, N.~J.
\newblock 2013.
\newblock Bursting your (filter) bubble: strategies for promoting diverse
  exposure.
\newblock In {\em Proceedings of the 2013 conference on Computer supported
  cooperative work companion},  95--100.
\newblock ACM.

\bibitem[\protect\citeauthoryear{Shani and
  Gunawardana}{2011}]{shani2011evaluating}
Shani, G., and Gunawardana, A.
\newblock 2011.
\newblock Evaluating recommendation systems.
\newblock In {\em Recommender systems handbook}. Springer.
\newblock  257--297.

\bibitem[\protect\citeauthoryear{Su and Khoshgoftaar}{2009}]{su2009survey}
Su, X., and Khoshgoftaar, T.~M.
\newblock 2009.
\newblock A survey of collaborative filtering techniques.
\newblock {\em Advances in artificial intelligence} 2009:4.

\bibitem[\protect\citeauthoryear{Tak{\'a}cs and
  Tikk}{2012}]{takacs2012alternating}
Tak{\'a}cs, G., and Tikk, D.
\newblock 2012.
\newblock Alternating least squares for personalized ranking.
\newblock In {\em Proceedings of the sixth ACM conference on Recommender
  systems},  83--90.
\newblock ACM.

\end{thebibliography}
